\documentclass[12pt, preprint]{aastex}

\shorttitle{ An \ion{H}{1}  study of three shell-type supernova remnants 
in the local arm region}
\shortauthors{YAR UYANIKER, UYANIKER \& KOTHES}

\begin{document}

\title{Distance of three Supernova Remnants from \ion{H}{1} line observations
in a complex region: G114.3+0.3, G116.5+1.1, and  CTB~1 (G116.9+0.2)}

\author{AYL\.IN YAR UYANIKER\altaffilmark{1,2}, 
        B\"ULENT  UYANIKER\altaffilmark{1,2},
        ROLAND  KOTHES\altaffilmark{1,3} }

\altaffiltext{1}{National Research Council, 
       Herzberg Institute of Astrophysics, 
       Dominion Radio Astrophysical Observatory, 
       P.O. Box 248, Penticton, B.C., 
       V2A 6J9 Canada}
\altaffiltext{2}{Max-Planck-Institut f\"ur Radioastronomie, 
                Auf dem H\"ugel 69,
                D-53121, Bonn, Germany}
\altaffiltext{3}{
         Department of Physics and Astronomy, 
         The University of Calgary, 
         2500 University Dr. NW, 
         Calgary, AB, 
         T2N 1N4 Canada}

\email{aylin@mpifr-bonn.mpg.de, uyaniker@mpifr-bonn.mpg.de,
roland.kothes@nrc-cnrc.gc.ca}

\begin{abstract}

We present new radio continuum and \ion{H}{1} images towards the supernova 
remnants (SNRs) G114.3$+$0.3, G116.5$+$1.1, and G116.9$+$0.2 (CTB~1) 
taken from the Canadian Galactic
Plane Survey (CGPS). We discuss the dynamics of their 
\ion{H}{1} environment and a possible relationship of these SNRs with each 
other. We discovered patches of \ion{H}{1} emission surrounding G114.3$+$0.3 
indicating a location in the Local arm at a distance of about 700~pc in 
contrast to previous publications which proposed a Perseus arm location. 
The other two SNRs have radial velocities of $-$17~km s$^{-1}$ (G116.5+1.1) and 
$-$27~~km s$^{-1}$ (CTB~1) according to related \ion{H}{1}. 
However, the structure 
of the \ion{H}{1} and its dynamics in velocity space suggest a possible 
relation between them, placing both remnants at a distance of about 1.6~kpc. 
CTB~1 appears to be embedded in an \ion{H}{1} feature which is moving as a 
whole towards us with a velocity of about 10~km s$^{-1}$. Furthermore, the 
off-centered location of CTB~1 in a large \ion{H}{1} bubble indicates that 
the so-called breakout region of the remnant is in fact due to its expansion 
towards the low density interior of this bubble. 
We believe that the progenitor star of CTB~1 was an early B or O-type
star shaping its environment with a strong stellar wind in
which case it exploded in a Ib or Ic event.

\end{abstract}


\section{Introduction}

Towards Galactic Longitude $\ell\approx 115\degr$, there are three extended, 
shell-type supernova remnants within an area of a few degrees; G114.3+0.3,
G116.5+1.1, and G116.9+0.2 also known as CTB~1. The proximity of these
sources and their relative sizes suggest a possible association.
Kinematic distance estimates which are based
on the correlation with surrounding atomic material are very difficult due 
to poor resolution of available data. \ion{H}{1} absorption measurements
fail because the sources are not bright enough to be absorbed by the
warm foreground gas and possibly related
CO was not detected. As a result distances to these remnants are poorly
determined, especially for CTB~1, where the suggested values differ
by more than a factor of two. In an attempt to improve the distance
estimates of these remnants and to learn more about their environment
we analyzed new \ion{H}{1} observations from the CGPS 
\citep{tay03}, which provides data at an unprecedented resolution 
of about 1\arcmin.

In this paper, we also present new radio continuum and polarization 
images of the
region at 1420 MHz, from the CGPS database. A multi-frequency study of
these remnants, especially of CTB~1, at 408, 2700, 4850 and 10500 MHz,
including a comparative polarization analysis will be the topic of
another publication (Yar-Uyan{\i}ker et al. 2003, in preparation).

In what follows we will review the characteristics of each individual
source.

\subsection{G114.3+0.3}
G114.3$+$0.3 was identified as a supernova remnant by \citet{rei81}
based on a comparison of radio continuum data at 11~cm and 21~cm. They
suggested a distance of 3.4~kpc based on possibly related \ion{H}{1} at
a systemic velocity between $-$35~km/s and $-$45~km/s. \citet{fur93} and 
\citet{kul93} associated this remnant with the pulsar PSR~2334+61. The 
distance of the pulsar was obtained from its dispersion measure of
$DM=55\pm10$ pc cm$^{-3}$ -- now a more accurate value is available
which gives DM$=58.38\pm 0.09$ \citep{mit03} -- as 2.4(2.5)~kpc, using the 
model of free
electrons by \citet{tay93}. The newly calculated model for the 
Galactic distribution of free electrons by \citet{cor02} is unfortunately
not very useful for us since the authors incorporated  a 
void in the direction of G114.3$+$0.3 most likely to accommodate for the larger
distance to the SNR predicted by previous authors. As
already discussed by \citet{mit03}, the rotation measure (RM) and
dispersion measure of pulsars increase anomalously in the
presence of small scale \ion{H}{2} regions along the line of sight.  
They found that PSR 2334+61 has an anomalous RM behaviour and the calculated
foreground magnetic field along the line of sight is more than twice that
of pulsars which are supposed to have no foreground object affecting
their DM and RM values.
Thus, the distance determination from the DM or RM of the pulsar may not be
very accurate. 

G114.3+0.3 has also been detected in optical emission \citep{fes97, 
mav02}. Given the fact that optical emission suffers strongly from
interstellar absorption the existence of optical emission from this
SNR points to a rather close distance.

\subsection{G116.5+1.1}
G116.5+1.1 was also discovered
and identified by \citet{rei81} at 21~cm and 11~cm wavelengths.  Using
low resolution data, from the Maryland-Green Bank survey \citep{wes82},
\citet{rei81} noted a depression in \ion{H}{1} between $-45$ km s$^{-1}$ and 
$-60$ km s$^{-1}$, which they translated to a distance between 3.6~kpc 
and 5.2~kpc. They did not notice any other feature in their data which 
might be associated with this SNR.  \citet{fes97} have detected faint
optical filaments in H$\alpha$ and \ion{S}{2}. The most prominent
one, about $30\arcmin$ long, positionally coincides with the bright
radio rim. 

\subsection{CTB~1 (G116.9+0.2)}
CTB~1 has been extensively studied in radio \citep{wil73, dic80,
rei81, lan82} and in optical emission \citep{van73, loz80,
fes97}. H$\alpha$
observations give a mean optical velocity of $v_{\rm LSR}=-33\pm2$ km
s$^{-1}$ \citep{loz80}.  \citet{rei81} found a systemic velocity of
$-$35 km s$^{-1}$ to $-50$ km s$^{-1}$, by relating the SNR to an \ion{H}{1}
structure partially encompassing the remnant from the south. However the
resolution of their data ($\sim10\arcmin$) does not allow a conclusive
agreement between the remnant's shell and the \ion{H}{1} filaments.
They also mention the existence of an expanding atomic
hydrogen shell associated with CTB~1, whose parts can be traced in the
red and blue shifted components at velocities $v_{\rm LSR}=-20$ km s$^{-1}$ to
$-30$ km s$^{-1}$ and $v_{\rm LSR}=-80$ km s$^{-1}$. However, the full
extent of this shell, expanding with a velocity of 20~km~s$^{-1}$ to 
40~km~s$^{-1}$, remained undetermined due to the confusion with the local
and outer arm emission.

\citet{lan82} used \ion{H}{1} data at 5$\arcmin$ angular resolution,
achieving a velocity resolution of $\sim2.6$ km s$^{-1}$.  Three major
components in their spectra are evident, signaling the Local arm gas at
$-$20 to $+$20 km s$^{-1}$, the Perseus arm gas at $-$30 to $-$70 km
s$^{-1}$ and the Outer arm gas at $-$90 to $-$110 km s$^{-1}$.  They
did not find an exact correlation at any velocity.  However,
they found an area of \ion{H}{1} emission between $-$15 km s$^{-1}$
and $-$35 km s$^{-1}$ which peaks at $-30$ km s$^{-1}$ and associated
the remnant with this area.  This velocity is also comparable to that
obtained from the H$\alpha$ data and implies a distance of 2.8~kpc
using a flat rotation curve with $R_\odot=8.5$ kpc and $v_\odot=220$
km s$^{-1}$ \citep{kerr86}. Absorption measurements towards
the western rim of CTB~1 provided no distance, because the foreground
\ion{H}{1} brightness ($\sim25$ K T$_{\rm B}$) is too high in
comparison to the continuum emission ($\sim4$ K T$_{\rm
B}$ with respect to the background).  By comparing nearby open clusters, O associations,
and \ion{H}{1} regions \citet{lan82} have concluded that CTB~1 must
have a distance of $2.0\pm0.4$ kpc.  This would place the
SNR in the interarm region between Local arm and Perseus arm.

According to \citet{fic86} all three SNRs belong to a group of
objects located within a large \ion{H}{1} shell at a velocity of
$v_{\rm LSR}=-46$ km s$^{-1}$ , which translates to a distance of 4.2
kpc using a flat rotation curve. \citet{fic86}, on the other hand,
quote the corresponding distance as 3.4 kpc, by comparing the velocity
of the center of the Perseus arm with the distances of the \ion{H}{2}
regions in the neighboring area.

As demonstrated above, determining the distance of a supernova remnant
is a difficult task.  The established distances to many remnants are
based on a relation between the surface brightness of the SNR to its
diameter, usually called the $\Sigma-D$ relation. Since it has been
shown that such a relation does not exist or just gives upper 
distance limits \citep[see][for a thorough discussion]{gre84, ber86, ber87},
we ignore previous distance estimates of these SNRs from the 
$\Sigma-D$ relations and
concentrate on more direct observational evidence to determine
distances; such as morphological and circumstantial coincidence of
continuum and \ion{H}{1} data, \ion{H}{1} absorption features and
relation of SNRs with the surrounding astronomical objects with known
distances.

\section{Observations}\label{obs}
Radio continuum and \ion{H}{1} line observations at 1420 MHz were
carried out simultaneously with the Synthesis Telescope of the
Dominion Radio Astrophysical Observatory (DRAO) \citep{lan00} as part
of the CGPS. The angular resolution of the final data products at this
frequency is $49\arcsec \times 49\arcsec~{\rm{cosec}} ({\delta})$ for
the radio continuum and $59\arcsec \times
59\arcsec~{\rm{cosec}}({\delta})$ for the \ion{H}{1} data.  The
\ion{H}{1} data have a velocity resolution of 1.3 km s$^{-1}$
covering 256 channels from $-$158 km s$^{-1}$ to $+$52 km s$^{-1}$
with a channel separation of 0.82~km s$^{-1}$. The rms noise
in the images at full resolution is 0.03 K T$_{\rm B}$ in continuum
and 3  K T$_{\rm B}$ in \ion{H}{1}. A detailed
description of the data processing routines can be found in
\cite{wil99}. Except the polarization maps, single antenna data are incorporated into
the synthesis maps to ensure accurate representation of all structures 
up to the largest scales. The low spatial frequency \ion{H}{1} data are from the
Low Resolution DRAO Survey of the CGPS region observed with the DRAO
26-m Telescope \citep{hig00}.

\section{Results}

\subsection{Radio continuum emission}

In this section we briefly describe the structure of the radio continuum 
emission in the area around the three SNRs. All of the SNRs display 
partial shell structures
placed on top of smooth  emission plateaus. Fig.~\ref{c21}
displays a large region of approximately $5\degr\times3\degr$ in size, which
includes all of the SNRs discussed in this paper and shows their
relative positions to each other.  The contour map representations of
these remnants, shown in Fig.~\ref{con2} and \ref{con1}, provide details
about the surrounding emission plateaus.  

CTB~1, at lower latitude in
Fig.~\ref{con1}, is the smallest with an angular diameter of 
$38\arcmin\times 32\arcmin$ and has the highest surface brightness.  
Its emission is
concentrated in two bright arcs in the south-east and in the west;
and these two arcs, apart from a region with less emission (the
breakout region) towards north-east, form an almost circular
shape. 

G114.3+0.3 (Fig.~\ref{con2}) shows a dominant thin arc in the
south-west. Within  its boundary is the bright \ion{H}{2} region S~165,
which is not related to the supernova remnant \citep{rei81}. 
The angular size of this SNR is
$82\arcmin\times65\arcmin$. 
The associated pulsar PSR 2334+61 is located
at the center of the remnant \citep{fur93, kul93, bec96}.

G116.5+1.1 (Fig.~\ref{con1}) is located just north of  CTB~1,
and has an angular size of $ 81\arcmin\times 62\arcmin$. 
The strongest emission from the remnant
is concentrated in a relatively thick arc, $\sim18\arcmin$ wide, at
the opposite side of CTB~1.  A smooth emission plateau with a brightness
of 0.5 K T$_{\rm B}$, which is encircling the two SNRs, is placed
on top of the smooth Galactic background component
of about $5$ K T$_{\rm B}$, in this direction. 
This emission plateau might indicate  a possible 
association.

\subsection{Polarized emission}
Polarization properties of the DRAO Synthesis Telescope are
given in \citet{tay03} and preparation of polarized survey images
are discussed in detail by \citet{uya03}. Portions of the polarization
data relevant to this paper, showing all three remnants, are plotted
in Fig.~\ref{pol} at  2$\arcmin$ resolution.  
Polarized emission above 0.1 K T$_{\rm B}$ is detected from both
G114.3 and G116.5.  They appear as  partial shells with
fine structure in polarization. CTB~1, on the other hand, appears
to be weaker. 

In the following sections we will use the polarization data as an
argument to claim a close distance for the SNRs. According to
\citet{uya03} there exists a maximum distance from which the polarized
emission  reaches us without being totally depolarized.
This distance depends on the electron density, the strength of the 
magnetic field and the
distance  from which the emission  originates. \cite{uya03} called
this distance  the ``{\em polarization horizon}''. Clearly the horizon is not
single valued and varies with the wavelength of the observation and
physical parameters along the line of sight. This, however, implies a
strong relationship between the detectability of the 
polarized emission and the distance of the object.

A phase shift of 90$\degr$ between two polarization vectors would
cause total depolarization. To achieve this phase shift a rotation
measure of 36 rad m$^{-2}$ at 21~cm is required. Towards the SNRs the magnetic
field strength parallel to the line of sight is B$_{\|}=1 ~\mu$G \citep{mit03}.
This yields a dispersion measure of 44 cm$^{-3}$ pc according to 
Eqn.~\ref{eqn1}, for which the electron density model of \citet{tay93}
gives a distance of 2.2 kpc in this direction.
This value is similar to the one found by \citet{uya03} towards
$82^\circ < \ell < 95^\circ$ and implies that rotation measure 
sets an upper limit on the path length of the observable polarized
emission, namely the distance of the object.  

\subsection{Neutral Hydrogen}

The 21 cm line data from the CGPS reveal new information about the
kinematics of the neutral hydrogen surrounding these three SNRs.  
Figures~\ref{h2} and \ref{h1} present  averaged \ion{H}{1} maps towards 
these SNRs.  

Bright and large \ion{H}{1} emission patches surround 
G114.3+0.3, centered at a velocity of $v_{\rm LSR}= -6.5$ km s$^{ -1}$
(see Fig.~\ref{h2}). No other \ion{H}{1} feature which shows any
resemblance to the SNR can be found in the entire \ion{H}{1} data cube.
Using a flat rotation model for the Milky Way with 
$R_\odot=8.5$~kpc and $v_\odot=220$~km s$^{-1}$ this would translate
to a kinematic distance of 700~pc, which is significantly closer than all
previously published distance estimates all of which proposed a location 
within Perseus arm.
The structure of the \ion{H}{1} emission does not indicate that
the environment was shaped by the stellar wind of the
progenitor star, in which case we would expect a more dynamic
environment dominated by thin filaments.

The \ion{H}{1} environment of G116.5+1.1 as displayed in Fig.~\ref{h1} (left
panel) looks like it was carved out by the expanding shockwave of the
supernova explosion. To the
west and south there are bright patches of \ion{H}{1} following nicely the
curvature of the radio bright shell. To the east most of the area shows a lack
of \ion{H}{1}. Like for G114.3+0.3 it looks like the ambient medium was
not influenced by stellar wind effects of the progenitor star. 
The central radial velocity of the \ion{H}{1} 
structure is about $-17$~km s$^{-1}$ which translates to a kinematic distance
of about 1.6~kpc.

CTB~1, on the other hand, is surrounded by an ``open-end-ring'' of neutral
hydrogen (see Fig.~\ref{h1}, right panel) centered at 
$v_{\rm LSR}= -27$ km s$^{-1}$, which translates to a
kinematic distance of 2.5 kpc. The shape of the 
surrounding \ion{H}{1} ring coincides well with the breakout appearance 
of the SNR's radio continuum emission as seen in Fig.~\ref{con1}. In this
case the structure of the surrounding \ion{H}{1} looks more dynamic.
Here we find thin filaments and
shells indicating a highly dynamic environment in contrast to the smooth
bright patches surrounding the other SNRs (better presented in
Figs~\ref{ch1}-\ref{ch2}). The progenitor star of CTB~1
obviously had a strong stellar wind and affected the ambient medium
significantly.

In the area around CTB~1 the local arm and the Perseus arm are
very distinct in velocity space as clearly demonstrated in
the velocity-distance diagram in Fig.~\ref{veldis}, where
we plotted SNRs and \ion{H}{2} regions within 20\degr of Galactic longitude 
of the SNRs as a function of distance to emphasize this fact. 
Local objects have radial
velocities above $-20$ km~s$^{-1}$ while Perseus arm sources show
radial velocities below $-30$ km~s$^{-1}$. The decision whether G114.3 and
G116.5 reside in the local arm or the Perseus arm, based on the
radial velocities of $-6.5$ km s$^{-1}$ and $-17$ km s$^{-1}$ respectively, is
made very simple. However, the associated \ion{H}{1} shell
of CTB~1 with a radial velocity of $-27$ km~s$^{-1}$ indicates an
interarm location close to Perseus arm.  Because of the complex and highly
dynamic environment of the SNR, which suggests a highly active
progenitor star with a strong stellar wind, we believe that CTB~1 must 
reside in one of the spiral arms.  This makes CTB~1 either a blueshifted
local or a slightly redshifted Perseus arm object. 

\section{Discussion}

\subsection{G114.3+0.3}

Besides the possibly related \ion{H}{1} structure we found, a fact that
most of the time is somewhat subjective anyway, we have many other observational
and circumstantial evidence which can give us constraints on the distance
of this SNR. First of all, optical filaments were detected by \citet{fes97} 
and \citet{mav02}. Since optical emission suffers from interstellar
absorption its presence indicates a rather low distance. In the direction
of G114.3+0.3 this would mean that a Perseus arm location is rather
unlikely. This fact is supported by the presence of polarized emission
in the CGPS at 21~cm (see Fig.~\ref{pol}) which indicates a distance in front
of the so-called polarization horizon, which was found to be at about 2~kpc
for CGPS data \citep{uya03}. These two constraints already imply 
a location in the Local arm rather than the Perseus arm. An additional
upper limit can be derived from the distance to S~165 which is not
depolarizing the 11~cm continuum emission at its line of sight and hence
must be located behind the SNR \citep{rei81}. A new spectroscopic distance
estimate \citep{bran93} puts this \ion{H}{2} region at a distance of 1.6~kpc.
These additional characteristics agree with the earlier assessment that
G114.3+0.3 is related to patches of \ion{H}{1} at a distance of 700~pc.
At this distance the SNR would be $18 \times 11$~pc in size.

At this low distance the dispersion measure and rotation measure of
PSR~2334+61 become another point of debate. From the model by \citet{tay93}
we get a foreground dispersion measure of DM$_{\rm fore}= 13$~cm$^{-3}$pc. But the
pulsar has a dispersion measure of about $58$~cm$^{-3}$pc, which would give
a distance of 2.5~kpc. On the other hand we also find a very high
rotation measure of $-100$~rad~m$^{-2}$ which combined with the dispersion
measure leads to a foreground magnetic field parallel to the line
of sight of B$_\parallel = -2.1$~$\mu$G;  
which is significantly higher than the expected value of
B$_\parallel = -1$~$\mu$G \citep{mit03}. The enhanced magnetic field
indicates an additional structure along the line of sight with a strong
magnetic field and a high electron density causing the additional dispersion
and rotation measure. Since the pulsar is located within a supernova remnant
the nature of this additional structure should be clear. Hence we now want
to investigate the possibility that the shell of the expanding SNR itself
causes the high RM and DM values. 

The rotation measure RM is defined by:

\begin{equation}
\frac{\rm RM}{\rm B_\parallel} = 0.82\cdot {\rm DM} = 0.82\cdot \int {\rm n}_{\rm e} \cdot {\rm dl},
\label{eqn1}
\end{equation}
here n$_{\rm e}$ represents the electron density in [cm$^{-3}$] and l the
pathlength in [pc]. Assuming a foreground dispersion measure of 
DM$_{\rm fore} = 13$~cm$^{-3}$~pc and a mean foreground magnetic field
along the line of sight of B$_\parallel = -1$~$\mu$G, we get internal values
of DM$_{\rm SNR} = 45$~cm$^{-3}$~pc, RM$_{\rm SNR} = -89$~rad~m$^{-2}$, and
B$_\parallel^{\rm SNR} = -2.4$~$\mu$G, for the SNR. 

The ratio between the internal and external magnetic field strengths
results in a compression of 2.4. Assuming the same compression ratio for
the interstellar medium, swept up by the expanding shockwave and a radius
of 7 pc gives a width of 1.2 pc for the shell. Together with the internal
dispersion measure we get an electron density of 38~cm$^{-3}$ within the
shell. Assuming the swept up material is fully ionized, a Hydrogen to Helium
ratio of 9:1, and again a compression ratio of 2.4 yield an average ambient
density of about 14 atoms~cm$^{-3}$.  These are reasonable values and strengthen
the proposed nearby location.

The surrounding patches of \ion{H}{1} indicate a more homogeneous
ambient medium for the supernova explosion rather than a stellar wind
structured one. This would indicate a B2 or later type progenitor star 
-- since these are not massive enough to create a strong stellar wind --
and a type II supernova event.

At 700~pc distance we get a mean radius of about 7~pc for the SNR. 
Following \citet{cio88} a supernova remnant with an explosion
energy of E$_0$~[$10^{51}$~erg] and an ambient medium density of 
n$_0$~[cm$^{-3}$] would enter the so-called pressure driven snowplow
phase (PDS) at a radius R$_{\rm PDS}$~[pc] defined by:
 
\begin{equation}
{\rm R}_{\rm PDS} = 14.0\cdot \frac{{\rm E}_0^{2/7}}{{\rm n}_0^{3/7}},
\end{equation}
for solar abundances. For an ambient density of n$_0=14$~cm$^{-3}$ and
an explosion energy of E$_0=10^{51}$~erg we derive R$_{\rm PDS} = 4.5$~pc.
This would indicate that G114.3+0.3 has already entered the PDS phase and
indeed it would require an explosion energy of E$_0\approx 5\cdot 10^{51}$~erg
for the SNR to be still in the so-called energy conserving Sedov phase. 
With the above parameters we can calculate the
mass of the swept up material to be M$_{\rm sw} = 480$~M$_\odot$ and from 
the equations determined
by \citet{cio88} we can also estimate an age of about 7700~yrs for G114.3+0.3.

\subsection{G116.5+1.1}

As for G114.3+0.3 we can get additional constraints on the distance from
the presence of observable optical emission \citep{fes97} and polarized
1.4~GHz radio continuum emission in the CGPS (see Fig.~\ref{pol}). 
This indicates a Local arm
location, most likely closer than 2.2~kpc. The \ion{H}{1} structure we found,
surrounding the SNR, has a radial velocity of about $-17$~km~s$^{-1}$ which
translates to a location within the Local arm at about 1.6~kpc. Similar to
G114.3+0.3 the surrounding \ion{H}{1} consists of diffuse and smooth emission
patches which do not display any dynamics. It looks more like the expanding
shockwave of the supernova explosion created the hole, the SNR resides in,
by taking away the material in the interior. Hence as for G114.3+0.3 we
believe the progenitor star was a B2 or later type star and exploded in a 
type II event. The diffuse emission to the east in contrast to the sharp
outer boundary to the west indicates that the star exploded inside this
\ion{H}{1} cloud but to the east the shockwave already left the cloud
in a kind of breakout giving the SNR its smooth structure in that direction.

At a distance of 1.6~kpc the SNR has a mean radius of 16.5~pc. Given the
similarity in the emission structure to G114.3+0.3 we might assume that  
G116.5+1.1 has also just entered the PDS phase. Assuming an explosion energy
of $10^{51}$~erg we get with equation (2) a lower limit for the ambient
density of 0.8~cm$^{-3}$. This indicates that the SNR has indeed entered
the PDS phase because a lower density would be quite unusual for a location
within a spiral arm. Only within stellar wind bubbles we would expect
a lower density but there is no evidence for such a structure around 
G116.5+1.1. For an ambient density of 1~cm$^{-3}$ we would get an age of about
15000~yrs and for 5~cm$^{-3}$ about 50000~yrs. The mass of the material 
swept up by the expanding shockwave would be 450~M$_\odot$ and 
2250~M$_\odot$ respectively. Since the SNR is still rather
bright in radio continuum emission a smaller age implying an earlier
evolutionary state is to be prefered.

\subsection{CTB~1 (G116.9+0.2)}

The associated \ion{H}{1} shell
of CTB~1 with a radial velocity of $-27$ km s$^{-1}$ indicates an
interarm location. But because of the complex and highly dynamic
environment of the SNR we believe that it must reside in one
of the spiral arms: in the interarm one expects to have less material
in which a remnant expands easier and more or less symmetrically. 
This makes CTB~1 either a blueshifted local or a redshifted Perseus arm 
object. To make a decision
we have to carefully investigate the behavior and dynamics of surrounding
structures in the neutral hydrogen going from CTB~1 to the
local and Perseus arm (Fig.~\ref{ch1} and.Fig.~\ref{ch2}) to find a 
connection to either of them.

Starting from low negative velocities in the local arm a smooth
and large \ion{H}{1} bubble forms around CTB~1, best pronounced
between $-16$ km s$^{-1}$ and $-18$ km s$^{-1}$. This bubble was 
most likely created
by stellar wind effects and/or supernova explosions a long time
ago. The diffuse structure indicates a late stage of evolution
in which the structure starts to merge with the surrounding medium.
Towards higher negative velocities the edges of the bubble slowly
transform into bright thin filaments which are moving closer to
the radio bright shell of CTB~1 with decreasing velocity. Bright
filamentary structures indicate a highly dynamic environment in
an earlier stage of evolution where energetic interaction is
still taking place. Between $-25$ km s$^{-1}$ and $-27$ km s$^{-1}$ 
those filaments
create a bubble around CTB~1 which is closest around the radio bright
shell and open to the other side where we see only smooth low
surface brightness radio emission. This structure is best pronounced at
$-27$ km s$^{-1}$. Going from there to even higher negative velocities
the structure of the bubble remains constant while it slowly fades
away. At the same time smooth emission from the Perseus arm is
starting to move in from the south. There appears to be 
a connection between local arm gas and the bubble around
CTB~1, but not between CTB~1 and the Perseus arm. Hence CTB~1
is a blue-shifted local object.

The most likely scenario to explain the observed structures and
the velocity shift is that the progenitor star of CTB~1 resided in
the smooth bubble visible at local arm velocities. The star
was not located in the center but in the south-west and closer to
us. Thus the stellar wind and later the supernova shock wave
impacted on the edges of the bubble first in the south-west giving
the SNR its half shell shape. Since the progenitor of CTB~1
was located closer to us relative to the center of the bubble it
also hits the edges closer to us giving it a shift in velocity
since this part of the bubble is moving towards us. Hence the systemic
velocity of the \ion{H}{1} structure around CTB~1 is the same
as for the smooth bubble at $-17$ km s$^{-1}$, assuming of course that
this bubble is not shifted as well. The local arm position of CTB~1
is also supported by the presence of polarized emission, although
it is very weak.

This systemic velocity translates to a distance of
1.6 kpc. We note that both SNRs G116.5+1.1 and CTB~1 appear to be located 
in the same large \ion{H}{1} complex at a distance of $\sim$1.6 kpc and
the breakout region of G116.5+1.1 might even be located inside the
same bubble as CTB~1. Since the structure around CTB~1 has a radial
velocity of about $-27$~km s$^{-1}$ it is moving towards us at a velocity
of $\sim10$ km s$^{-1}$ relative to the main body of the complex. 
At a distance of 1.6~kpc the linear size of the
remnant would be $18\times15$~pc. 

With this interpretation, we can also understand the emission
structure of the breakout region of CTB~1 in the radio continuum map. If the
remnant is located in the south-west of the bubble rather than in its
center, its shockwave arrives at the lower edge of the \ion{H}{1}
bubble earlier creating the bright radio shell and giving rise to its
incomplete circular shape. Hence the north-east opening of the remnant
is not necessarily due to a breakout, 
but simply the result of a freely expanding
shockwave which is traveling into the low density interior of the
bubble.

\section{Summary}

We have presented new sensitive radio continuum and \ion{H}{1} data at
1$\arcmin$ resolution towards the three supernova remnants G114.3+0.3,
G116.5+1.1 and CTB~1 (G116.9+0.2). The strength of the \ion{H}{1} analysis is
the combination of high resolution data with large spatial
coverage. G116.5+1.1 and CTB~1 appear to be located
within the same \ion{H}{1} complex.  Based on the surrounding \ion{H}{1} 
structure and dynamics we derived new distance estimates for these SNRs.
We also concluded that G114.3+0.3 and G116.5+1.1 are expanding inside an
area not much effected by stellar wind effects of their progenitors implying
supernova explosions of type II with B2 or later type progenitor stars. 
CTB~1, however, is expanding in a highly dynamic environment indicating 
an early B or O-type progenitor star exploding in a Ib or Ic event.
We also believe that G114.3+0.3 and G116.5+1.1 are expanding inside an
area not much affected by stellar wind effects of their progenitors in which
case the supernova explosions would have been type II events with a B2
or later type progenitor star.

According to our analysis G114.3+0.3 is the closest remnant among the
three SNRs with a distance of about 700~pc which gives this object
a mean diameter of about 14~pc. The SNR is in the so-called
pressure driven snowplow phase, has swept up about 480~M$_\odot$
of material and is probably around 7700~yrs old. The other two remnants are
located at a distance of 1.6 kpc. This gives G116.5+1.1 an average diameter
of 33~pc, making this SNR the biggest of the three discussed here. It has
also reached the PDS phase and with typical SNR parameters we get an age
more than twice as much as G114.3+0.3 and probably more than 450~M$_\odot$
of swept up material.

CTB~1 appears to be located at the
edge of a large \ion{H}{1} bubble which expands with a velocity of
$\sim10$ km s$^{-1}$ towards us with respect to the systemic velocity
of the whole system ($-17$ km s$^{-1}$). The distance of 1.6~kpc gives
the remnant a size of about 16.5~pc. We argue that the
``opening'' of CTB~1 in the northeast is not an effect of a breakout
but simply the result of expanding into the interior of its surrounding
bubble. 

\acknowledgments{ The Dominion Radio Astrophysical Observatory is a
National Facility operated by the National Research Council.  The
Canadian Galactic Plane Survey is a Canadian project with
international partners, and is supported by the Natural Sciences and
Engineering Research Council (NSERC). We wish to thank Tom Landecker 
and Wolfgang Reich for careful reading of the manuscript and discussions.}

\newpage

\begin{figure}]\centering
\caption{Radio continuum image of the three supernova remnants, CTB~1 
(G116.9+0.2), G116.5+1.1 and G114.3+0.3 from the CGPS at 1420 MHz.}
\label{c21}
\end{figure}

\begin{figure}\centering
\caption{Total intensity map of G114.3+0.3 at 21 cm radio continuum.
Contours start at 5.5 K T$_{\rm B}$ total intensity up to 7.4 K T$_{\rm
B}$ in steps of 0.16 K T$_{\rm B}$.  The position of the PSR2334+61 is
indicated with a star. Towards G114.3+0.3 there are three \ion{H}{2}
regions, S163 ($\ell=113.5\degr, b=-0.6\degr$), S165
($\ell=114.6\degr, b=+0.2\degr$), and S166 ($\ell=114.6\degr,
b=-0.8\degr$).}
\label{con2}
\end{figure}

\begin{figure}\centering
\caption{Contour map of CTB~1 and G116.5+1.1 at 21 cm radio continuum.
Contours represent total intensity starting at 5.5 K T$_{\rm B}$ in
steps of 0.12 K T$_{\rm B}$ up to 6.7 K T$_{\rm B}$, in black. The
higher emission regions are represented by white contours starting at
7 K T$_{\rm B}$ and 7.5 to 10.5 K T$_{B}$ in steps of 1.0 K T$_{\rm
B}$.}
\label{con1}
\end{figure}

\begin{figure}\centering
\caption{Polarized intensity images towards the three SNRs at 2 arcminute
resolution.
Gray scale extends from 0.02 (white) to 0.15 K T$_{\rm B}$ (black). Contours 
start from 0.1 K T$_{\rm B}$ and run in steps of 0.025 K T$_{\rm B}$. 
White contours are overlaid to show the total intensity emission.}
\label{pol}
\end{figure}

\begin{figure}\centering
\caption{Averaged \ion{H}{1} map towards the supernova remnant
G114.3+0.3.  The \ion{H}{1} line emission is averaged between
$v_{\rm LSR} = -4.8$ km s$^{-1}$ and $v_{\rm LSR}= -7.2$ km s$^{-1}$. The
\ion{H}{1} intensity contours (in black) start from 40 K T$_{\rm B}$
up to 80 K T$_{\rm B}$ with 10 K T$_{\rm B}$ increasing steps.  The 21
cm radio continuum emission is represented by white contours at
from 6.2 K T$_{\rm B}$ and  6.6 K T$_{\rm B}$.
The noise in the maps is 1.7 K T$_{\rm B}$, due to averaging
of images.}
\label{h2}
\end{figure}

\begin{figure}\centering
\caption{ {\em Left:} Averaged \ion{H}{1} maps towards the SNRs
CTB~1 and G116.5+1.1. The left-hand panel shows the \ion{H}{1} line
emission averaged over velocity channels between $v_{\rm LSR} =
-15.5$ km s$^{-1}$ and $ v_{\rm LSR}= -18.5$ km s$^{-1}$.  
The \ion{H}{1} contours
(in black) start from 10 K T$_{\rm B}$ and run to 38 K T$_{\rm B}$, in
7.0 K T$_{\rm B}$ steps. {\em Right:} \ion{H}{1} intensities
averaged from $v_{\rm LSR} = -27.5$ to $ v_{\rm LSR}= -30.5$ km
s$^{-1}$. Contour levels are 6 to 46 K T$_{\rm B}$ with steps of 10 K
T$_{\rm B}$.  The 21 cm radio continuum emission, in both images, is
overlaid as white contours at  6.0 and 7.0 K T$_{\rm B}$.
The noise in the maps is 1.5 K T$_{\rm B}$, due to averaging
of images over 3 km s$^{-1}$.}
\label{h1}
\end{figure}

\begin{figure}\centering
\caption{ The radial velocity of SNRs and  \ion{H}{2} regions within 
$20\degr$ in Galactic longitude of CTB~1 as a function of distance. The 
radial velocities of our three target supernova remnants are indicated. 
Information about the \ion{H}{2}  regions were taken from \citet{bran93}.
The radial velocities and distances of the SNRs were taken from: 
\citet{kot01} (G106.3+2.7), \citet{kot02} (CTB~109), \citet{goss88}\&
\citet{reed95} (Cas~A), \citet{pine93} (CTA~1), \citet{reyn99} \& \citet{chev80} (Tycho), Yar-Uyan{\i}ker et al.~(2004; in preparation) (G126.2+1.6 and G127.2+0.5), \citet{gre82} \& 
\citet{wall94} (3C58), and \citet{rout91} (HB~3).} 
\label{veldis}
\end{figure}

\begin{figure}\centering
\caption{Distribution of the \ion{H}{1} emission observed towards the
supernova remnants CTB~1 and G116.5+1.1, as channel maps.  The radio
continuum emission from CTB~1 shown  by  a contour
at 7 K T$_{\rm B}$  to indicate the position of the SNR on the \ion{H}{1} channel maps.}
\label{ch1}
\end{figure}

\begin{figure}\centering
\caption{Same as Fig.~\ref{ch1} but for different velocity intervals as labeled on the
panels.}
\label{ch2}
\end{figure}

\end{document}